\documentclass{article}

 \newcommand{\bitem}{\begin{itemize}}
 \newcommand{\eitem}{\end{itemize}}

\usepackage[figuresright]{rotating}
\RequirePackage{rotating}
\RequirePackage{multirow}
\RequirePackage{algpseudocode,algorithm,algorithmicx}
\RequirePackage{subcaption}
\RequirePackage{natbib}
\usepackage{amsmath}
 \usepackage[flushleft]{threeparttable}
\usepackage{longtable}
\usepackage{booktabs}
\usepackage{caption}
\usepackage{url}
\usepackage[all]{xy}

\usepackage{amsfonts}
\usepackage{amssymb}
\usepackage{amsthm}

\theoremstyle{lemma}
\newtheorem{lemma}{Lemma}
\theoremstyle{theorem}
\newtheorem{theorem}[lemma]{Theorem}
\newtheorem{remark}[lemma]{Remark}

\title
      {SuRF: a New Method for Sparse Variable Selection, with Application in
  Microbiome Data Analysis}

\author{Lihui Liu$^{1,*}$, 
Hong Gu$^{1}$, Johan Van Limbergen$^{2}$ and Toby
Kenney$^{1}$\\
$^{1}$Department of Mathematics and Statistics,  Dalhousie University, Halifax, Canada\\
$^{2}$Department of Pediatrics, Dalhousie University, Halifax, Canada}

\begin{document}




\label{firstpage}


\maketitle

\begin{abstract}
In this paper, we present a new variable selection method for
regression and classification purposes. Our method, called {\bf
  Su}bsampling {\bf R}anking {\bf F}orward selection (SuRF), is based
on LASSO penalised regression, subsampling and forward-selection
methods. SuRF offers major advantages over existing variable selection
methods in terms of both sparsity of selected models and model
inference.  We provide an \texttt{R} package that can implement our
method for generalized linear models. We apply our method to
classification problems from microbiome data, using a novel
agglomeration approach to deal with the special tree-like correlation
structure of the variables. Existing methods arbitrarily choose a
taxonomic level {\em a priori} before performing the analysis, whereas
by combining SuRF with these aggregated variables, we are able to
identify the key biomarkers at the appropriate taxonomic level, as
suggested by the data. We present simulations in multiple sparse
settings to demonstrate that our approach performs better than several
other popularly used existing approaches in recovering the true
variables. We apply SuRF to two microbiome data sets: one about
prediction of pouchitis and another for identifying samples from two
healthy individuals. We find that SuRF can provide a better or
comparable prediction with other methods while controlling the false
positive rate of variable selection.
\end{abstract}

\subsection*{keywords}
Variable selection; Generalised Linear Models; LASSO; SuRF;
Microbiome; Forward selection; Stability selection; Identifying biomarkers.


\section{Introduction \label{section intro}}

Traditional statistical methods face lots of challenges when analyzing
high-dimensional data. These challenges occur in model fitting,
variable selection and model diagnosis. A series of regularized models
from the field of data mining have become popular inference approaches
for high dimensional data such as gene expression. The most well-known
methods include LASSO regression \citep*{tibshirani1996regression}, the
elastic-net regression model
\citep*{zou2005regularization} and various variations such as group
LASSO~\citep*{groupLASSO}, Bayesian LASSO~\citep*{BayesianLASSO}, etc. LASSO is based on penalising the model by
the sum of the absolute values of coefficients of all variables and hence it is a soft thresholding method so that some variables are eliminated due to a resulting zero coefficient. This has the
advantage of selecting sparse models. In addition, it is a
suitable method to use for tree-structured data, such as microbiome
data, as we discuss in Section~\ref{TreeStructure}.

There are a few issues with the use of LASSO for microbiome data. The
first is inference --- LASSO can select a parsimonious model, but it
does not provide a direct quantitative assessment of the significance
of each variable selected. For scientific and clinical research, it is
vital to include these assessments of the significance of variables
($p$-values). There is a method related to this
matter~\citep*{lockhart2014significance} but, in practice, high
dimensional data rarely satisfies the weak collinearity assumption
needed. The more robust approach of~\citet*{TibshiraniTaylor} is only
available for Gaussian response variables. Secondly, LASSO provides
only a list of variables, with coefficients, but in many cases very
strong correlation exists between some variables, either of which
might be selected with no indication that the other variables might
have an almost equally strong association with the response
variable. The choice of which variables are selected can be very
unstable.

We are particularly interested in applications to microbiome
research. The microbiome is the collection of all bacteria present in
a location, e.g. a person's gut, and one of the main
questions of microbiome research is the relationship between phenotypes
(e.g., healthy versus disease groups) and the microbiome.
The data consist of counts of various types of microbes, classified
into Operational Taxonomic Units (OTUs). Due to the vast number of
OTUs and the cost of samples, the sample size is always much
smaller than the number of OTUs.

We introduce a variable selection method, SuRF, based on regularized
regression and subsampling of observations in the generalized linear
model setting. This method provides a $p$-value for each variable, and gives information on the stability of the
selected variables.

There has been previous work on dealing with the lack of stability in
LASSO, such as \citet*{zakharov2013stable} and
\citet*{grave2011trace}. A promising recent approach to this issue
which has some similarity to our SuRF method is stability
selection~\citep*{meinshausen2010stability}.  We show in our
simulations that SuRF gives much better performance in variable
selection than stability selection.

\section{{\bf  Su}bsampling {\bf R}anking {\bf F}orward selection (SuRF)}

The framework of SuRF has two main steps: creating a list of ordered
predictors that have been selected most frequently by LASSO on
subsamples of the observations, and selecting variables by forward
selection on the list of ordered variables. The second step consists
of determining the significance of the variables in terms of
likelihood ratio statistics via sequential permutation tests and
implementing ANOVA with forward selection to eliminate or alleviate
the issue of surrogate variables. The algorithm pseudo code is
provided in Appendix A of the web supplementary material.

\subsection{Variable Ranking}\label{ranking}

The subsampling approach plays an important role in formulating the
list of top predictors. This technique is widely used in many recent
methods for variable selection and the details were summarised well
by~\cite*{dezeure2015high}. In each of the splits, a part of the data
is used to select variables by LASSO. We focus on LASSO here because
of the tree structure in microbiome data and its empirically better
screening performance than other regularized
models~\citep*{buhlmann2014high}; however, the subsampling could be
applied with other variable selection methods. We recommend about 90\%
of the data for this purpose when the sample size is extremely limited
but otherwise the proportion should make a minimal difference in
results (see Web Table~11). We also recommend taking stratified subsamples (subsamples
having the same proportion in each class as the true data) in classification problems since if
the subsamples are not balanced, this can affect prior probabilities
of each class, resulting in worse classification.  The sample
splitting is repeated in a similar manner a large number, $B$, of
times. Some literature suggests that $B=50$ or $B=100$ is
sufficient~\citep*{dezeure2015high}, but there is little cost to using
larger $B$ to ensure better results. For each subsample, we record the
variables selected by LASSO, with the tuning parameter selected by
cross-validation over the subsample. We rank variables by the
frequency they are selected over the $B$ subsamples. (Ties are broken
by reduction of deviance residuals from models containing all
higher-ranked variables.)  The order of variables can be interpreted
as measuring the strength of the association with the response variable.
 
\subsection{Sequential permutation tests with ANOVA forward selection}

The forward selection involves sequentially testing the null
hypothesis that no variables beyond the currently selected variables
are good predictors for the response variable, given the currently
selected variables. In forward selection, the order in which variables
are added to the model can be important. At each stage, if multiple
predictors are significant, we add the one ranked highest by our
variable ranking procedure. Because we are testing multiple predictors
at each stage, the log-likelihood ratio statistic does not follow a
$\chi^2$ distribution. The predictors are not independent, so
Bonferroni correction cannot be used. Instead we calculate the critical
value empirically using permutations.

More specifically, we start with a list of candidate variables,
containing all variables in the order found using the ranking method
from Section~\ref{ranking}, and a list of selected variables being
initially empty. At each step, we generate a random permutation for
all observations, and apply it only to the variables in the candidate
variable list. This breaks the relationship between the candidate
variables and the response variable, but preserves the correlation
structure among the candidate variables and between the response
variable and the selected variables. Now for each candidate variable,
we compute the log-likelihood ratio statistic between the current
model and the model with this variable added. We record the largest
log-likelihood ratio statistic. We repeat this process for many more
permutations (we usually use 200 permutations as a compromise between
accuracy and speed), to obtain the null distribution of the maximum
log-likelihood ratio statistic. We use the $1-\alpha$ $ (\alpha=0.05)$
percentile of this null distribution as our critical value, denoted as
$D^i_{1-\alpha}$ for the $ith$ variable in this forward
sequential variable selection procedure.  We now return to the
original unpermuted data, and for each candidate variable in the
ranking order, we calculate the log-likelihood ratio statistic between
the current model, and the model with this candidate variable
added. We select the first candidate variable for which this statistic
exceeds $D^i_{1-\alpha}$, and add this variable to the model (and
remove it from the candidate variable list). We then generate a new
distribution, with new permutations and repeat the same
procedure. When the log-likelihood ratio statistic for each candidate
variable is no greater than the critical value, we terminate the algorithm
and output the current model as the final model selected.

For each variable added to the model, SuRF has computed a $p$-value
based on the comparison of the likelihood ratio statistic with the
empirically calculated null distribution. This $p$-value is based on
the increase in predictive ability over the variables that were
already included in the model. That is, the $p$-value is for the null
hypothesis that all true variables are in the current model.  A
variable which is a surrogate for a variable that has been already
selected may not have a significant $p$-value if there is not
significant evidence that this variable improves prediction over the
surrogate already selected.

\subsection{Theory}

SuRF is designed to combine the best parts of three methods: stability
selection, LASSO and forward selection. The advantages and
disadvantages of these methods are as follows.
%
  Forward Selection provides clear $p$-values at each stage, but 
heavily depends on the order of variables 
entering the model.
LASSO does well at identifying the correct
variables, but does not provide $p$-values and often
selects too many variables.
  Stability selection is robust to outliers,
but can ``fall between two stools'' with
surrogates, and offers only limited $p$-values.


Because the final selection in SuRF is using forward selection, we
will base our theory on this approach. We want to show that
asymptotically, SuRF will select the true variables. This occurs in
two stages. Firstly, the true variables must be ranked highly by the
subsampling procedure. This relies on the performance of LASSO at
identifying the true variables. It is known that LASSO is consistent
provided the irrepresentability condition holds \citep*{ZhaoYu}.  In
this case, the subsampling is guaranteed to select the true variables
before other variables, which overcomes the danger with forward
selection that the surrogate will be selected first, preventing the
true variables from entering the model.  In addition, the subsampling
approach offers some robustness, allowing the top variables to be
highly ranked even if some outliers might make them less highly ranked
in some subsamples.

Secondly, assuming the true variables have been ranked above other
variables, the hypothesis test must reject the null model. We show
that, asymptotically, even for many cases with $p\gg n$, this will
happen using the following theorems, which are proved in Appendix B of
the online supplementary materials.


\begin{theorem}\label{TrueVariablesPass}
Suppose we have a true model ${\mathbb E}(Y|X)=g^{-1}(X\beta)$, where
$Y|X$ follows an exponential family distribution, $g$ is a link
function, and $\beta_i$ is zero for all but a fixed number $p_{\textrm{true}}$ of
predictors. Suppose further that we have already selected the set $S$
of variables, all of which satisfy $\beta_j\ne 0$ and that there is at
least one variable $i\not\in S$ with $\beta_i\ne 0$. Let $D(k)$ be the
log-likelihood ratio statistic for the variable $k$. Then there is
some $j\not\in S$ with $\beta_j\ne 0$ such that $D(j)=O(n)$, where $n$
is the number of observations.
\end{theorem}


\begin{theorem}\label{NullDistributionBound}
Suppose $X$ is an $n\times p$ matrix and $Y$ is a random vector
(independent of $X$) from the exponential family distribution fitted
by the model , then the maximum deviance of a single column of $X$ as
a predictor of $Y$ has survival function bounded by $S(x)\leqslant
\frac{2pe^{-\frac{x}{2}}}{\sqrt{2\pi x}}$.
\end{theorem}

From Theorem~\ref{NullDistributionBound}, the critical value at level
$\alpha$ is bounded above by the solution to
$\frac{2pe^{-\frac{x}{2}}}{\sqrt{2\pi x}}=\alpha$, which for all
reasonable $p$ and $\alpha$ satisfies $x\leqslant
-2\log\left(\frac{\alpha}{2p}\right)$. On the other hand,
Theorem~\ref{TrueVariablesPass} says the test statistic is
asymptotically $O(n)$ for at least one true variable. Therefore,
provided $\frac{\log(p)}{n}\rightarrow 0$, or equivalently
$p=o\left(e^{\epsilon n}\right)$ for any $\epsilon>0$, the true
variables must be selected.

\begin{remark}
  Note that the bound on the critical value in
  Theorem~\ref{NullDistributionBound} is the critical value for a
  Bonferroni correction for the multiple testing. We use the empirical
  value in practice because the Bonferroni critical value is too
  large, since it does not account for correlation between
  predictors.
\end{remark}

\subsection{Tree Structure}\label{TreeStructure}

SuRF can be applied to any exponential-family GLM variable selection
problem. However, our application of interest is microbiome data. In
this section we discuss the particular way we have adapted SuRF to
deal with such data.

Microbiome data typically consist of proportions of OTUs present in
each sample. OTUs are clusters of DNA sequences, usually clustered at
97\% similarity, approximately equivalent to species-level
resolution. We are working with a GLM $g({\mathbb
  E}(Y|X))=\beta_0+X\beta$, where $X$ is the column-centralised OTU
data matrix with each column representing an OTU variable. The
phylogenetic relationships among OTUs provide us with prior knowledge
about $\beta$. Namely, we expect the $\beta_i$ to be close for
closely-related OTUs because of phenotypic similarity. We reflect this
prior knowledge via the regularisation of the coefficients. We choose
to base this on the taxonomic tree, rather than more detailed
phylogenetic trees, because estimation of the taxonomic tree is more
robust, and the taxonomic tree is easily available from the output of
most pipelines. However it is trivial to use a phylogenetic tree
instead.

A common practice is to aggregate variables at an arbitrarily chosen
taxonomic level, usually genus or phylum. That is, to replace
the original data matrix $X$ by the aggregated data matrix
$\widetilde{X}=XC$, where $C$ is the clustering matrix at the chosen
level. For example, 
$$C_{ij}=\left\{\begin{array}{ll}1&\textrm{if OTU }i\textrm{ is in
  phylum }j,\\0&\textrm{otherwise.}\end{array}\right.$$
Now, fitting a model $g({\mathbb E}(Y|X))=\widetilde{X}\alpha+\beta_0$
is equivalent to fitting $g({\mathbb E}(Y|X))=X\beta+\beta_0$, where
$\beta=C\alpha$. That is, this regularisation consists of the
restriction $\beta=C\alpha$, namely that OTUs from the same phylum
have the same coefficients.

While aggregating at a sufficiently high taxonomic level can have the
convenient consequence that classical statistical methods can be
applied, the aggregated data may lack the resolution to answer the
scientific questions, or may lead us to make unsupported or false
generalisations. On the other hand, the large noise when analysing at
a low taxonomic level may obscure general patterns, and not provide a
satisfactory prediction \citep*{hou2015classification}.

The trouble with aggregation at a certain taxonomic level is that it
converts the soft prior expectation that coefficients for OTUs in the
same group should be similar into a hard requirement that the
coefficients be equal, even if this is disproved by the data. Instead,
we penalise the extent to which the coefficients differ. More
formally, instead of setting $\widetilde{X}=XC$, we set
$\widetilde{X}=X(C,I)$, where $(C, I)$ is a matrix whose first columns
are $C$, and whose remaining columns are the identity matrix. There
are now multiple ways to represent a given model in terms of the
variables in $\widetilde{X}$ because of the linear dependence between
columns. The regularisation means that only a single way to represent
the model minimises the penalty. For any coefficient vector $\beta$ in
the model $g({\mathbb E}(Y|X))=X\beta+\beta_0$, many vectors $\alpha$
satisfy $\beta=(C, I)\alpha$. These vectors can all be considered
correct models. However they are distinguished by the penalty term
$\lVert\alpha\rVert_1$ (since the loglikelihood is the same based on
either $\beta$ or $\alpha$).  It can be shown that the penalty term is
minimised when for any $j$ at the higher taxonomic level, $\alpha_j$
is the median of $\{0\}\cup\{\beta_i|C_{ij}=1\}$, and for any $i$ in
cluster $j$ (i.e. $C_{ij}=1$) $\alpha_i=\beta_i-\alpha_j$. We can
apply the same method after constructing similar aggregations at every
taxonomic level. The resulting penalty for a particular coefficient
vector $\beta$ is the most parsimonious total change of coefficients
over the taxonomic tree structure. We clarify this with an example:

\begin{figure}

\caption{Example tree}\label{treeconcepts}

\hfil \begin{subfigure}{0.46\textwidth}

\caption{Variables}
 
   \xymatrix@C=1em@R=2ex{&&&0\ar@{-}[d]\\
    &&& X_{11}\ar@{-}[dl] \ar@{-}[dr] \\
    & &X_{10}\ar@{-}[dl]&& X_9\ar@{-}[d] \ar@{-}[ddr] \\
    & X_7 \ar@{-}[dl] \ar@{-}[dr] \ar@{-}[d] &&& X_8\ar@{-}[dl] \ar@{-}[d] &\\
  X_1 & X_2 & X_3 & X_4 & X_5 & X_6 & \\
  }
  \end{subfigure}%

\vskip\baselineskip
\hfil \begin{subfigure}{0.46\textwidth}
\caption{Coefficients $\beta$}
  \xymatrix@C=1em@R=2ex{&&&0\ar@{-}[d]^0\\  
  &&& 0\ar@{-}[ddll]_1 \ar@{-}[dr]^{-0.5} \\
  &&&& -0.5\ar@{-}[d]_{-0.5} \ar@{-}[ddr]^0 \\
  & 1\ar@{-}[dl]_0 \ar@{-}[dr]^{1.5} \ar@{-}[d]^1 &&& -1 \ar@{-}[dl]_{-1} \ar@{-}[d]^0 \\
 1 & 2 & 2.5 & -2 & -1 & -0.5  \\  
 }

\end{subfigure}
\hfil\begin{subfigure}{0.46\textwidth}
\caption{Coefficients $\beta'$}
  \xymatrix@C=1em@R=2ex{&&&0\ar@{-}[d]^0\\  
  &&& 0\ar@{-}[ddll]_2 \ar@{-}[dr]^{-0.5} \\
  &&&& -0.5\ar@{-}[d]_{-0.5} \ar@{-}[ddr]^0 \\
  & 2\ar@{-}[dl]_0 \ar@{-}[dr]^{0.5} \ar@{-}[d]^0 &&& -1 \ar@{-}[dl]_{-1} \ar@{-}[d]^0 \\
 2 & 2 & 2.5 & -2 & -1 & -0.5  \\  
 }

\end{subfigure}

(a) shows a taxonomic tree relating the OTU variables
$X_1,\ldots,X_6$. We create additional variables $X_7,\ldots,X_{10}$
by aggregating the variables below. Let $X$ be the original OTU data
matrix $\left(X_1\cdots X_6\right)$. In (b), we consider the estimate
$Y=X\beta+\beta_0$ where $\beta=(1,2,2.5,-2,-1,-0.5)^T$ and in (c), we
consider the estimate $Y=X\beta'+\beta_0'$ where
$\beta'=(2,2,2.5,-2,-1,-0.5)^T$.

The coefficients in the expanded model are shown on the branches of
the trees, and the values shown at internal nodes are cumulative sums
of the branches above. For leaf nodes, these are the coefficient
$\beta$ in the original model.
  
\end{figure}
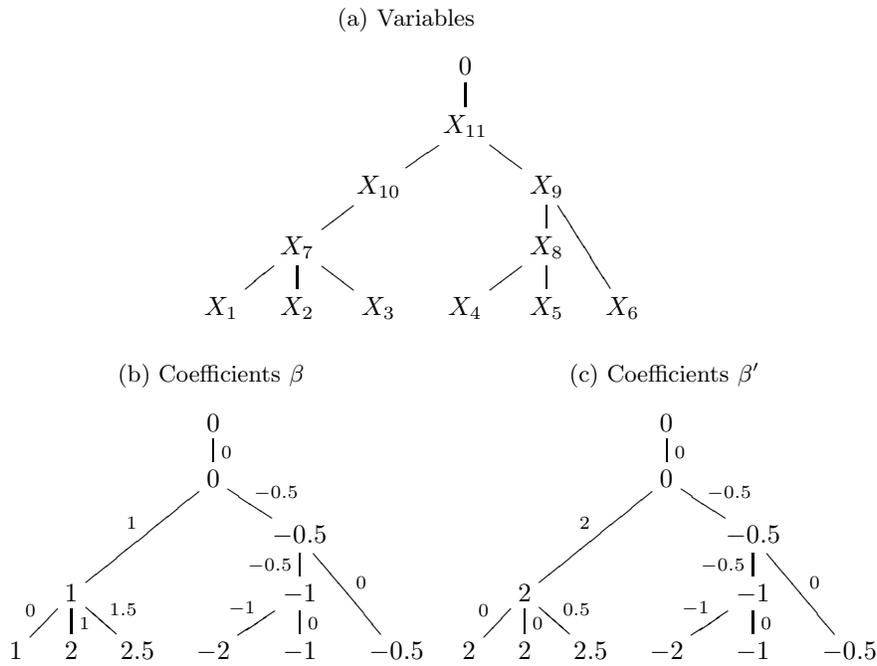

Figure~\ref{treeconcepts}(a) shows a small taxonomic tree containing
OTUs $X_1,\ldots,X_6$. We create the combinations $X_7=X_1+X_2+X_3$, $X_8=X_4+X_5$,
$X_9=X_4+X_5+X_6$ and $X_{11}=X_1+\cdots+X_6$. We do not consider
the combination $X_{10}$, because it is equal to $X_{7}$.
For the coefficient vector
$\beta=(1,2,2.5,-2,-1,-0.5)^T$,
i.e. the model $g({\mathbb E}(Y|X))=X_1+2X_2+2.5X_3-2X_4-X_5-0.5X_6$, the most parsimonious
coefficients in terms of the expanded set of predictors are $\alpha=(0,1,1.5,-1,0,0,1,-0.5,-0.5,0,0)$ as shown in
Figure~\ref{treeconcepts}(b). That is, 
$g({\mathbb E}(Y|X))=X_2+1.5X_3-X_4+X_7-0.5X_8-0.5X_9$ is equivalent to the original
estimate, but is given a lower penalty by LASSO.
Similarly, for the coefficient vector $\beta'=(2,2,2.5,-2,-1,-0.5)^T$,
in Figure~\ref{treeconcepts}(c), the most parsimonious coefficients in
the aggregated model are
$\alpha'=(0,0,0.5,-1,0,0,2,-0.5,-0.5,0,0)$. In the original model, the
penalty assigned to $\beta'$ is
$\lambda(|2|+|2|+|2.5|+|-2|+|-1|+|-0.5|)=10\lambda$, which is
 larger than the penalty
 $\lambda(|1|+|2|+|2.5|+|-2|+|-1|+|-0.5|)=9\lambda$ assigned to
$\beta$, whereas, for the aggregated model, the penalty assigned to
$\alpha$ is $$\lambda(|0|+|1|+|0|+|1|+|1.5|+|-0.5|+|-0.5|+|0|+|-1|+|0|)=5.5\lambda$$
and the penalty for $\alpha'$ is
$$\lambda(|0|+|2|+|0|+|0|+|0.5|+|-0.5|+|-0.5|+|0|+|-1|+|0|)=4.5\lambda$$
so the penalty for $\alpha$ is larger. Thus, the aggregation approach
uses the same space of models, but a different regularisation, which
can affect the selected model. In this aggregated setting, we will
often say something like ``We select the higher level variable
$X_{7}$'' as a shorthand for ``We select a model in which variables
$X_1$, $X_2$ and $X_3$ are included, but constrained to have equal
coefficients.''

\citet{yan2018rare} are independently developing a similar method
involving adding aggregated variables to alter the
regularisation. Their approach is tailored to text mining
problems, and consequently differs from ours in a couple of respects:
Firstly, they include an additional penalty for the coefficients at
leaf nodes. This does not make sense for OTU data, since leaf nodes
are clusters of lower level strains, so should not be treated
differently. Furthermore, additional penalty for coefficients at leaf
nodes creates a new optimisation problem. Secondly, their method does
not scale the variables before regularisation. For LASSO,
standardisation makes predictors more comparable, so that penalties
are equivalent. For their count data, the counts are already
equivalent.  For our tree-based LASSO, it is less clear what
standardisation means. Further work on fine-tuning the procedure to
produce a better penalty that more accurately reflects this is outside
the scope of the current paper, which focusses on the SuRF procedure,
but is a topic the authors plan to address in future work.

The theory for this augmented version of LASSO is still not
developed. It is not possible to apply the standard theory for the
augmented set of predictors, because there are many representations of
the true model using the augmented predictors. Described in terms of
the augmented predictors, even the notion of consistency is
challenging to define --- there are multiple correct sets of selected
augmented variables, and when we convert back to the original
variables, it can be challenging to even determine whether or not a
given original variable has been selected. Developing a new theory
about the consistency of augmented LASSO is beyond the scope of the
current paper, but is an interesting area for future work.

There are several ad-hoc methods in the literature to incorporate 
tree structure into the LASSO model, for example \citet*{Xiao2018}. However, an
advantage of our aggregated LASSO method is that it is trivial to
also incorporate covariates which do not fit into the tree structure,
simply by not creating aggregated variables for them. This approach
can also be applied to multiple hierarchical clustering structures on
the same set of variables e.g. all clades from all gene trees for a
given data set. We can do this by simply adding a set of aggregated
variables for each clustering.

\section{Simulation}
\subsection{Study 1: Simulation using variables from higher taxonomic levels}

Our first simulation is based on the original microbiome data matrix
$X$ from the pouch data (afferent limb site)
in~\cite*{tyler2013characterization} (see details of the data in
Section~4.1). We examine our method under the null case (no variable
is significantly associated with the outcome variable) and under
various sparse settings using variables from higer taxonmical levels:
phylum or class. These settings were chosen to be similar to the
results from the real data analysis on that dataset. For each
simulated dataset we compare the performance of SuRF with several
existing popular variable selection methods: LASSO,
VSURF~\citep*{genuer2015vsurf} and stability selection. VSURF uses the
variable importance from the random forest method to select
variables. Stability selection performs LASSO variable selection on a
large number of subsamples of the data, and selects the variables that
are selected by LASSO for a large proportion of these subsamples.

Due to the
sample size, we cannot afford to hold out a test sample, so only
in-sample prediction (same predictor matrices for the training and
test data) results are available.  The penalty parameter $\lambda$ for
LASSO is obtained by a 5-fold cross validation procedure and we use
the $\lambda$ which gives an error within 1 standard error of the best
model. For stability selection, we adopt a range of threshold
probabilities recommended in \citet*{meinshausen2010stability},
between 0.6 and 0.9 (results for stability with cutoff 0.6 and 0.9 in
simulation study 1 and 2 are shown in the Tables~\ref{tab1}--\ref{tab3}
and results with cutoffs 0.7 and 0.8 in web tables~5--7), and use the
default family error rate upper bound parameter of 1.  VSURF offers
variable selection for different objectives: interpretation and
prediction. We compare only the variables selected for prediction,
which are always a subset of the variables selected for
interpretation.

For assessment of results, we look at both the variables selected and
the predictive accuracy. Variable selection can be used either for
interpretation or for prediction. For microbiome data, the
interpretation can be challenging because of the large number of
surrogate variables. We therefore view predictive accuracy as the
primary objective of our variable selection, with interpretation a
secondary goal. However, selection of the true variables is important
for both prediction and interpretation. Therefore, we have included it
in the results of our simulations.

\subsubsection{Null case}
We simulated 200 datasets with binary outcomes randomly generated from
a Bernoulli distribution with probability 0.3. In this case, a good
variable selection algorithm is the one that selects no variables in
most cases and on average includes the least number of noise
variables. The mean and standard deviation of number of noise
variables selected are summarized in Table~\ref{tab1}(b).

\subsubsection{Sparse setting}
We simulate four scenarios for the true predictors: a single variable (Bacteroidetes) that has a strong
surrogate variable (S1 in Table~\ref{tab1}); a single variable (Firmicutes) that has no extreme
surrogate variable (S2 in Table~\ref{tab1}); two unrelated variables (Bacteroidetes and
Firmicutes) with equal signal strength (S3 in Table~\ref{tab1}); and two variables
(Bacilli and Clostridia) that together make up the majority of a
single phylum (S4 in Table~\ref{tab1}); The same data matrix $X$ is used to simulate the inflammation
outcome at different signal-to-noise ratio (SNR) levels under a
logistic regression model. The overall SNR is approximately $\frac{\mathop{\textrm{Var}}(P(Y))}{E(P(Y)(1-P(Y)))}$.

In the simulation, the $\beta$ coefficient(s) (see Table~\ref{SimulationCoeffs}) are
chosen so that SNR is approximately 0.7 (weak), 1 (Fair), and 3
(strong). The coefficients that achieve these SNRs depend on a number
of factors, such as the distribution and correlation of the
predictors. Because OTU abundance is often heavy-tailed, and the
response is Bernoulli, these coefficients can be large compared with
other datasets. For example, in Scenario~1, the coefficient for
Bacteroidetes is $-2.40$ in the low SNR case, $-2.84$ in the medium
SNR case, and $-4.58$ in the high SNR case. In all cases, the
intercept of the model is set to be the same. The coefficients of
Bacteroidetes and Firmicutes are set to be negative and positive
respectively according to the relationship from the original data.

\begin{table}[!htbp]
  \centering
  \caption{Coefficients for four different simulation scenarios}\label{SimulationCoeffs}
\resizebox{\linewidth}{!}{%
    \begin{tabular}{c|r|rrrr}
\hline
    \multirow{2}{*}{Case} &  \multirow{2}{*}{SNR} & \multicolumn{4}{c}{Coefficients of variables ($\beta$)} \\
                  &       & Bacteroidetes & Firmicutes &  Bacilli     &Clostridia  \\
          \hline
  \multirow{3}{8em}{Single variable with one strong surrogate (Case 1)} & High  & -4.58 &       &       &  \\
           & Fair  & -2.84 &       &       &  \\
         & Low   & -2.40  &       &       &  \\
\hline
    \multirow{3}{8em}{Single variable with no extreme surrogate (Case 2)}  & High  &       & 5.00     &       &  \\
         & Fair  &       & 2.32  &       &  \\
          & Low   &       & 1.85  &       &  \\
\hline
    \multirow{3}{8em}{Two variables with equal strength (Case 3)} &  High  & -4.87 & 4.39  &       &  \\
         & Fair & -1.82  & 1.64   &       &  \\
           & Low   & -1.42 & 1.28  &       &  \\
\hline
   \multirow{3}{8em}{Two variables equivalent to one variable (Case 4)} & High &   & &  4.76     &  4.76        \\
        & Fair &      & & 2.21      &   2.21      \\
       & Low   &&    &  1.75     &    1.75       \\
    \hline
   \end{tabular}}
\end{table}

The variable selection results for the null case and all four sparse
cases are given in Table~\ref{tab1}(a,b).  For Case~1, Bacteroidetes
is one of the major phyla existing in the human gut and this phylum is
mainly composed of the class Bacteroidia. The correlation of these two
variables in our data is almost one and they are deemed as a pair of
strong surrogate variables. Given such a high correlation, the
algorithm may identify either of them as the predictor and we deem
either case as correct. No variable selection method is able to
distinguish between them as predictors. However, we deem the selection
of both variables as the inclusion of a noise variable, since once the
first variable is included, the second variable does not give
additional information.

In Case 2 (S2 in Table~\ref{tab1}), Firmicutes is another dominant phylum
but there is no class that has a correlation with it as high as that
between Bacteroidetes and Bacteroidia. The class Clostridia has fairly
high correlation, but we deem Clostridia as an incorrect variable in
this case.

In Case 3 (S3 in Table~\ref{tab1}), the strengths of the two variables
(Bacteroidetes and Firmicutes) are set equal by adjusting the
coefficients according to their standard deviation. As in Case~1, we
deem Bacteroidia as a correct alternative to Bacteroides, but deem the
selection of both to constitute a noise variable.

In Case 4 (S4 in Table~\ref{tab1}), two class-level variables Bacilli and
Clostridia are the two true predictors, with the same
coefficients. The phylum Firmicutes is divided into three major
classes including Bacilli, Clostridia and Erysipelotrichi with the
former two classes representing the majority of the Firmicutes phylum
so that the sum of classes Bacilli and Clostridia is highly correlated
with total Firmicutes.  SuRF and other methods will often pick up
either the Firmicutes phylum as the true variable due to its surrogate
status, which is deemed as a correct result, or chose the combination
of the Firmicutes phylum and Erysipelotrichi class (or sometimes the
\textit{Incertae Sedis} genus which represents the majority of the
class Erysipelotrichi) as an equivalent linear combination to the sum
of classes Bacilli and Clostridia, which is an almost perfect
result. As the SNR decreases from high to low, preference is often
given to Firmicutes as the only true predictor.

The prediction results in these simulations are obtained by fitting a
logistic regression model on the selected variables for each
method. The results are shown in Table~\ref{tab1}(c).

Throughout the simulation study, SuRF performed well at eliminating
noise variables. Given the sequential nature of the variable
selection, the number of noise variables selected should be a
geometrically distributed variable with probability 0.95 (probability
that the selection process stops when all remaining variables are
noise variables) where we define the geometric random variable as
number of failures rather than number of trials, so the average number
of noise variables per data set should be $1/0.95-1=0.0526$. The
results from our simulation study are mostly consistent with this,
perhaps with slightly higher false positive rate on average.
Since the null hypothesis is that all true variables are in the model,
additional noise variables can be selected if they are ranked above
the true variables. Therefore, the reliability of the $p$-values is
better assessed by counting noise variables that are selected after
all true variables are already included in the model. This is
considered in more detail in Appendix~D.
Other methods
perform much worse in terms of false positives, with the exception of
stability selection with cut-off 0.9. Note that contrary to the claims
of \citet*{meinshausen2010stability}, the results of stability
selection are sensitive to the choice of cut-off probability.

Looking at false negatives, SuRF performs well in nearly all cases
compared with other methods. For cases with a single true variable, no
method has a significantly lower false negative rate than SuRF, not
even the methods whose false positive rate is an order of magnitude
higher than that of SuRF. For the two-variable cases, no method
identifies at least one of the true variables significantly more often
than SuRF. For Cases~3 and 4, LASSO selects two variables
significantly more often than SuRF (except in Case~3 with high
SNR). However, given the huge difference in number of noise variables
selected, it would be very surprising if LASSO did not select both
true variables significantly more often. The only method which had
comparable false positive rate to SuRF was stability selection with cut-off
0.9, and this performed much worse than SuRF in terms of false negatives.

From a prediction prospective, SuRF achieves very good in-sample
misclassification error rates compared to other methods. In Simulation
Study~2 (see Table~\ref{tab2}), we find that the other methods show
significantly better performance when assessed via in-sample test
error, compared with hold-out test data, whereas SuRF actually
performs worse on in-sample test data. If a similar pattern applies to
this study, then we would expect SuRF to outperform all other methods
on classification of test data. Performance on test data is more
important than on in-sample test data, because it better reflects the
usage --- in practice, we want a model that will generalise to new
data, which often will not have the same predictor values.

\begin{table}[htbp]
  \resizebox{0.9\textwidth}{!}{%
\begin{threeparttable}[t]

  \centering
  \caption{Simulation study 1}
    \begin{tabular}{r|l|rrrrr}
 \hline
    \multicolumn{1}{c|}{\multirow{2}[4]{*}{Senario }} &   \multicolumn{1}{c|}{\multirow{2}[4]{*}{SNR }}  & \multicolumn{1}{c}{\multirow{2}[4]{*}{\textbf{SuRF}}} & \multicolumn{2}{c}{\textbf{Stability}} & \multicolumn{1}{c}{\multirow{2}[4]{*}{\textbf{VSURF}}} & \multicolumn{1}{c}{\multirow{2}[4]{*}{\textbf{LASSO}}} \\
\cline{4-5}          
&       &       & \multicolumn{1}{c}{0.6} &  \multicolumn{1}{c}{0.9} &       &  \\
    \hline

  \multicolumn{1}{c}{} &    \multicolumn{1}{c}{}    &       &                     &       &       &  \\
\multicolumn{2}{}{}&\multicolumn{5}{c}{(a) \textbf{True positive results} \tnote{1}}\\
    \hline

    \multicolumn{1}{l|}{\multirow{3}[2]{*}{S1}} & High  & \multicolumn{1}{c}{100} & \multicolumn{1}{c}{98}  & \multicolumn{1}{c}{58} & \multicolumn{1}{c}{82} & \multicolumn{1}{c}{100} \\
          & Fair  & \multicolumn{1}{c}{98} & \multicolumn{1}{c}{88}  & \multicolumn{1}{c}{26} & \multicolumn{1}{c}{79} & \multicolumn{1}{c}{100} \\
          & Low   & \multicolumn{1}{c}{95} & \multicolumn{1}{c}{81} & \multicolumn{1}{c}{16} & \multicolumn{1}{c}{83} & \multicolumn{1}{c}{95} \\
    \hline
   
    \multicolumn{1}{l|}{\multirow{3}[1]{*}{S2}} & High  & \multicolumn{1}{c}{100} & \multicolumn{1}{c}{100} & \multicolumn{1}{c}{100} & \multicolumn{1}{c}{100} & \multicolumn{1}{c}{100} \\
          & Fair  & \multicolumn{1}{c}{100} & \multicolumn{1}{c}{100} & \multicolumn{1}{c}{93} & \multicolumn{1}{c}{93} & \multicolumn{1}{c}{95} \\
          & Low   & \multicolumn{1}{c}{97} & \multicolumn{1}{c}{99} & \multicolumn{1}{c}{71} & \multicolumn{1}{c}{83} & \multicolumn{1}{c}{87} \\
    \hline

    \multicolumn{1}{l|}{\multirow{3}[1]{*}{S3}} & High  & 100 (100) & 90 (100) & 24 (100) & 86 (100) & 100 (100) \\
          & Fair  & 66 (100) & 72 (99)  & 3 (63) & 63 (94) & 88 (99) \\
          & Low   & 35 (96) & 43 (92)  & 1 (46) & 49 (93) & 70 (98) \\
    \hline

    \multicolumn{1}{l|}{\multirow{3}[1]{*}{S4}} & High  & 19 (100) & 22 (100) & 1 (100) & 8 (100) & 57 (99) \\
          & Fair  & 9 (100) & 14 (100)  & 2 (62) & 16 (98) &  84 (84) \\
          & Low   & 8 (92) & 8 (98)  & 0 (32) & 5 (94) & 20 (66) \\
    \hline

 \multicolumn{1}{c}{} &   \multicolumn{1}{c}{}     &              &       &       &       &  \\
\multicolumn{1}{}{}&\multicolumn{6}{c}{(b) \textbf{False positive results:} average number of noise variables (SD)}\\
    
    \hline

 \multicolumn{1}{l|}{\multirow{1}[4]{*}{\textit{Null\tnote{2}} } }  &    &0.03 (0.18) &13.10 (0.67)  & 0.01 (0.07)&  3.96 (2.64)&  0.92 (5.15) \\ 

\hline

    \multicolumn{1}{l|}{\multirow{3}[2]{*}{S1}} & High  & \textbf{0.02} (0.14) & 4.06 (2.18)  & 0.46 (0.63) & 4.50 (3.11) & 22.45(21.62) \\
          & Fair  & \textbf{0.11} (0.35) & 1.78 (1.51) & 0.15 (0.46) & 4.76 (3.13) & 31.46(43.69) \\
          & Low   & 0.09 (0.32) & 1.24 (1.20) & \textbf{0.04} (0.20) & 4.58 (3.30) & 42.96 (55.03) \\
    \hline
     
    \multicolumn{1}{l|}{\multirow{3}[1]{*}{S2}} & High  & 0.06 (0.24) & 0.56 (1.09) & \textbf{0.00} (0.00) & 5.77 (2.97) & 24.04 (25.64) \\
          & Fair  & 0.11 (0.31) & 0.89 (1.16) &\textbf{0.08} (0.34) & 5.26 (2.87) & 33.92 (37.04) \\
          & Low   & 0.07 (0.26) & 0.93 (1.37) & \textbf{0.02} (0.14) & 5.00 (2.82) & 29.52 (42.46) \\
    \hline
       
    \multicolumn{1}{l|}{\multirow{3}[1]{*}{S3}} & High  & 0.05 (0.22) & 0.54 (0.81) & \textbf{0.01} (0.10) & 2.49 (2.27) & 18.79 (32.51) \\
          & Fair  & 0.06 (0.24) & 0.81 (1.04) &\textbf{0.04} (0.20) & 4.15 (2.76) & 31.57 (43.18) \\
          & Low   & 0.16 (0.40) & 1.12 (1.23)  &\textbf{0.04} (0.20) & 4.12 (2.98) & 27.61 (37.27) \\
    \hline

    \multicolumn{1}{l|}{\multirow{3}[1]{*}{S4}} & High  & 0.08 (0.31) & 0.84 (1.14)  & \textbf{0.03} (0.22) & 5.60 (2.85) & 26.24 (24.94) \\
          & Fair  & \textbf{0.11} (0.31) & 1.04 (1.45) & 0.14 (1.51) & 4.30 (2.52) & 19.56 (29.58) \\
          & Low   & 0.09 (0.29) & 0.82 (1.26) & \textbf{0.02} (0.14) & 4.33 (2.49) & 19.21 (33.25) \\
    \hline
 \multicolumn{1}{c}{} &  \multicolumn{1}{c}{}      &       &              &       &       &  \\
\multicolumn{1}{}{}&\multicolumn{6}{c}{(c) \textbf{In-sample}  average misclassficaition error rate (SD) }\\
    \hline
    \multicolumn{1}{l|}{\multirow{3}[2]{*}{S1}} & High  & \multicolumn{1}{l}{\textbf{0.095} (0.011)} & 0.103 (0.044) & 0.252 (0.194) & 0.108 (0.031) & 0.126 (0.062) \\
          & Fair  & \multicolumn{1}{l}{\textbf{0.190} (0.019)} & 0.219 (0.080) & 0.416 (0.141) & 0.240 (0.048) & 0.365 (0.142) \\
          & Low   & \multicolumn{1}{l}{\textbf{0.240} (0.082)}& 0.274 (0.101)\newline{} & 0.454 (0.107) & 0.276 (0.027) & 0.418 (0.120) \\
    \hline
    \multicolumn{1}{l|}{\multirow{3}[2]{*}{S2}} & High  & \multicolumn{1}{l}{0.093 (0.010)}  & 0.095 (0.011) & \textbf{0.092} (0.008) & 0.122 (0.018) & 0.224 (0.058) \\
          & Fair  & \multicolumn{1}{l}{ \textbf{0.173} (0.016) } & 0.178 (0.017)  & 0.187 (0.074) & 0.222 (0.023) & 0.294 (0.104) \\
          & Low   & \multicolumn{1}{l}{\textbf{0.210} (0.020)} & \textbf{0.210} (0.020)  & 0.282 (0.142) & 0.266 (0.024) & 0.368 (0.144)\\
    \hline
    \multicolumn{1}{l|}{\multirow{3}[2]{*}{S3}} & High  & \multicolumn{1}{l}{\textbf{0.102} (0.010)} & 0.115 (0.037)  & 0.196 (0.056) & 0.124 (0.015) & 0.228 (0.063) \\
          & Fair  & \multicolumn{1}{l}{0.204 (0.080)} &\textbf{0.192} (0.026)  & 0.316 (0.133) & 0.232(0.021) & 0.311 (0.100) \\
          & Low   & \multicolumn{1}{l}{0.262 (0.129)} &\textbf{0.232} (0.072)  & 0.365(0.139) & 0.265 (0.026) & 0.342 (0.127) \\
    \hline
    \multicolumn{1}{l|}{\multirow{3}[2]{*}{S4}} & High  & \multicolumn{1}{l}{\textbf{0.136} (0.030)}  & 0.139 (0.032)  & 0.152 (0.045) & 0.117 (0.018) & 0.204 (0.059) \\
          & Fair  & \multicolumn{1}{l}{\textbf{0.204} (0.016)} & 0.207 (0.012)  & 0.318 (0.147) & 0.220 (0.024) & 0.356 (0.160) \\
          & Low   & \multicolumn{1}{l}{ 0.245 (0.077) } & \textbf{0.231} (0.055) & 0.408  (0.129) & 0.254 (0.025) & 0.403 (0.152) \\
    \hline

    \end{tabular}%

  \label{tab1}%

\begin{tablenotes}
\item [1] In Scenarios~1 and 2, the table gives the total number of
  times the true single variable/surrogate variable is selected. In
  Scenario~3, the table gives the total number of two true variables
  selected and the number of times at least one of two true variables
  selected in the bracket. In Scenario~4, the table gives the number
  of times two true/surrogate variables are selected (perfect selection)
  and the numer of times the phylum Firmicutes is selected in brackets.
\item [2] The null Simulation is over 200 batches; all other senarios
  are 100 batches.\\
\end{tablenotes}

\end{threeparttable}%
  }
\end{table}

\subsection{Study 2: Simulation using variables from lower taxonomical levels}

Classification at the lower taxonomic level such as genus or species
level is more challenging \citep*{hou2015classification}. In this
section, we test the ability of SuRF to identify predictors at species
level. We base this simulation on the OTU abundance from the moving
picture data (see description in Section~4.2) simulating three
species-level true predictors. We simulate a binary response variable
at three levels of SNR, which are set using the same procedure as
described in Study~1. We also simulate a Gaussian response variable
with SNR levels set to 1 (Low), 3 (Fair), and 5 (High). For this case,
we fix the irreducible error at 1 for all SNRs and adjust signal
strength.  We compare methods on their true positive rate, false
postive rate, and either misclassification error rate (for binary
response) or median MSE and $R^2$ (for continuous response). In
Simulation study~1, we saw that stability selection had a much higher
false positive rate than SuRF. It is possible to control the false
positive rate more tightly in stability selection, by setting the
family error rate upper bound parameter. For Simulation~2, we attempt
to use this parameter to make the false positive rates more comparable
with SuRF by setting it equal to the average number of noise variables
selected by SuRF (i.e. the numbers in the SuRF column of Table~2(b)
and Table~3(b)).

The results are shown in Table~\ref{tab2} (binary response) and
Table~\ref{tab3} (continuous response). For both binary and Gaussian
response, SuRF can identify all three variables much more frequently
than any other method at all levels of SNR. Under a similar family
error rate, stability selection usually only identifies one true
variable. With this setting, the results of stability selection are
still sensitive to choice of cut-off probability, but less so than
with the default setting. For the binary outcome (see
Table~\ref{tab2}), SuRF gives the lowest misclassification error rate
for both in-sample data and test samples. For the continuous outcome
(see Table~\ref{tab3}), the median MSE for SuRF is almost identical to
the irreducible error of 1.  Other methods are much worse. We see the
same trend when we use $R^2$ as a performance measure.

\begin{table}[htbp]
 \resizebox{1\textwidth}{!}{
\begin{threeparttable}[t]
  \centering

  \caption{Simulation study 2 (Binary outcome)}
    \begin{tabular}{l|c|ccccccccc}
  \hline
    \multicolumn{1}{l|}{\multirow{2}[4]{*}{SNR}} & \multicolumn{1}{p{5.715em}|}{No of true } & \multicolumn{1}{c}{\multirow{2}[4]{*}{\textbf{SuRF}}} & \multicolumn{2}{c}{\textbf{Stability}} & \multicolumn{1}{c}{\multirow{2}[4]{*}{\textbf{VSURF}}} & \multicolumn{1}{c}{\multirow{2}[4]{*}{\textbf{LASSO}}} & \multicolumn{1}{c}{\multirow{2}[4]{*}{\textbf{RF}}} & \multicolumn{1}{c}{\multirow{2}[4]{*}{\textbf{SVM}}} \\
\cline{4-5}          & \multicolumn{1}{p{5em}|}{\parbox{5em}{variables\\ selected}} &       & 0.6    & 0.9   &       &       &       &  \\
   \hline
\multicolumn{1}{c}{}&\multicolumn{1}{c}{}&&&&&&&\\
\multicolumn{1}{c}{}&\multicolumn{8}{c}{\textbf{(a) Frequency of number of true varaibles selected over 100 simulations}}\\
    \hline
    \multicolumn{1}{l|}{\multirow{4}[2]{*}{High}} & 3     & \textbf{99}    & 0        & 0     & 20    & 30    & \multicolumn{2}{c}{\multirow{12}[6]{*}{N/A}} \\
          & 2     & 1     & 4       & 14    & 80    & 70    & \multicolumn{2}{c}{} \\
          & 1     & 0     & 96        & 86    & 0     & 0     & \multicolumn{2}{c}{} \\
          & 0     & 0       & 0     & 0     & 0     & 0     & \multicolumn{2}{c}{} \\
\cline{1-7}    \multicolumn{1}{l|}{\multirow{4}[2]{*}{Fair}} & 3     &\textbf{ 82 }  & 0         & 0     & 21    & 5     & \multicolumn{2}{c}{} \\
          & 2     & 18    & 3        & 3     & 78    & 90    & \multicolumn{2}{c}{} \\
          & 1     & 0     & 97     & 97    & 1     & 5     & \multicolumn{2}{c}{} \\
          & 0     & 0     & 0       & 0     & 0     & 0     & \multicolumn{2}{c}{} \\
\cline{1-7}    \multicolumn{1}{l|}{\multirow{4}[2]{*}{Low}} & 3     &\textbf{71}    & 0        & 0     & 9     & 4    & \multicolumn{2}{c}{} \\
          & 2     & 24    & 2     & 1     & 76    & 76    & \multicolumn{2}{c}{} \\
          & 1     & 5     & 97       & 97    & 15     & 20    & \multicolumn{2}{c}{} \\
          & 0     & 0     & 1         & 2     & 0     & 0     & \multicolumn{2}{c}{} \\
  \hline
\multicolumn{1}{c}{}&\multicolumn{1}{c}{}&&&&&&&\\
\multicolumn{1}{c}{}&\multicolumn{8}{c}{\textbf{(b) Number of noise varaibles selected over 100 simulations}}\\
    \hline
    \multicolumn{1}{p{4.215em}|}{High} & \multicolumn{1}{c|}{mean}& \textbf{0.120} & 0.76  & \textbf{0.120} & 8.71 & 68.93 & \multicolumn{2}{c}{\multirow{3}[2]{*}{}} \\
 \multicolumn{1}{p{4.215em}|}{} & \multicolumn{1}{c|}{SD}& (0.356) &  (0.452) & (0.327) &  (3.036) &(40.59) & \multicolumn{2}{c}{\multirow{3}[2]{*}{N/A}} \\
\cline{1-9}
    \multicolumn{1}{p{4.215em}|}{Fair} &\multicolumn{1}{c|}{mean}& 0.690& 0.340 & \textbf{0.02} & 7.080 & 62.45 & \multicolumn{2}{c}{} \\
\multicolumn{1}{p{4.215em}|}{} &\multicolumn{1}{c|}{SD}& (0.895) &  (0.476) & (0.141) & (3.183) & (46.98) & \multicolumn{2}{c}{} \\
\cline{1-9}
    \multicolumn{1}{p{4.215em}|}{Low}  &\multicolumn{1}{c|}{mean} & 0.610  &0.160&\textbf{0.010} & 6.590  & 61.92 & \multicolumn{2}{c}{} \\
  \multicolumn{1}{p{4.215em}|}{}  &\multicolumn{1}{c|}{SD} & (0.764) & (0.368)  &  (0.327) &  (0.100) &  (55.24) & \multicolumn{2}{c}{} \\
  \hline
\multicolumn{1}{c}{}&\multicolumn{1}{c}{}&&&&&&&\\
\multicolumn{1}{c}{}&\multicolumn{8}{c}{\textbf{(c) Mis-classfication error rate in test samples}}\\
   \hline
    \multicolumn{1}{p{4.215em}|}{High} &mean& \textbf{0.102}  & 0.383  & 0.412  & 0.288  & 0.290  & 0.292& 0.197 \\
\multicolumn{1}{p{4.215em}|}{} &SD&  (0.020) &  (0.033) &  (0.035) & (0.034) &  (0.041) &  (0.026) & (0.046) \\
  \hline
    \multicolumn{1}{p{4.215em}|}{Fair} &mean&\textbf{0.191}& 0.366 & 0.377 & 0.301& 0.323 & 0.291& 0.372 \\
 \multicolumn{1}{p{4.215em}|}{} &SD&  (0.028) & (0.034) &  (0.030) & (0.041) &(0.038) &  (0.032) & (0.080)\\
  \hline
   \multicolumn{1}{p{4.215em}|}{ Low}  &mean &\textbf{0.228}  & 0.361 & 0.366& 0.315 & 0.333  & 0.296& 0.390\\
 \multicolumn{1}{p{4.215em}|}{ }  &SD & (0.037) & (0.032)& (0.035) & (0.047) & (0.033) &(0.032) & (0.084)\\
   \hline
\multicolumn{1}{c}{}&\multicolumn{1}{c}{}&&&&&&&\\
\multicolumn{1}{c}{}&\multicolumn{8}{c}{\textbf{(d) In-sample mis-classfication error rate}}\\
   \hline
    \multicolumn{1}{p{4.215em}|}{High} &mean&\textbf{0.100} & 0.228& 0.295  & 0.126  & 0.169  & 0.126& 0.128 \\
\multicolumn{1}{p{4.215em}|}{} &SD&  (0.009) &  (0.038)  &  (0.033) & (0.011) &  (0.034) &  (0.011) & (0.014) \\
  \hline
    \multicolumn{1}{p{4.215em}|}{Fair} &mean&\textbf{0.220} & 0.309 & 0.350 & 0.259& 0.295 & 0.259& 0.277 \\
 \multicolumn{1}{p{4.215em}|}{} &SD&  (0.010) & (0.030) &  (0.010) & (0.017) &(0.029) &  (0.017) & (0.024)\\
  \hline
   \multicolumn{1}{p{4.215em}|}{ Low}  &mean &\textbf{0.259}   & 0.348& 0.371& 0.285 & 0.331  & 0.285&0.317 \\
 \multicolumn{1}{p{4.215em}|}{ }  &SD & (0.017) & (0.031)  & (0.012) & (0.020) & (0.034) &(0.020) & (0.032)\\
   \hline

    \end{tabular}%
  \label{tab2}%

\end{threeparttable}%
  }
\end{table}

\begin{table}[htbp]
 \resizebox{0.9\textwidth}{!}{
\begin{threeparttable}[t]
  \centering
  \caption{Simulation study 2 (Continuous outcome)}
    \begin{tabular}{l|c|cccccccc}
      \hline
    \multicolumn{1}{r|}{\multirow{2}[4]{*}{SNR}} & \multicolumn{1}{p{5.715em}|}{No of true } & \multicolumn{1}{c}{\multirow{2}[4]{*}{SuRF}} & \multicolumn{2}{c}{Stability} & \multicolumn{1}{c}{\multirow{2}[4]{*}{VSURF}} & \multicolumn{1}{c}{\multirow{2}[4]{*}{LASSO}} & \multicolumn{1}{c}{\multirow{2}[4]{*}{RF}}  \\
\cline{4-5}          & \multicolumn{1}{p{5.715em}|}{variables} &       & 0.6   & 0.9 &       &       &        \\
       \hline
\multicolumn{1}{c}{}&\multicolumn{1}{c}{}&&&&&&&&\\
\multicolumn{1}{c}{}&\multicolumn{9}{c}{\textbf{(a) Number of true variables selected}}\\
    \hline
    \multicolumn{1}{r|}{\multirow{4}[2]{*}{High}} & 3     &\textbf{ 99}    & 0     & 0         &   81    & 35    & \multicolumn{1}{c}{\multirow{12}[6]{*}{N/A}} \\
          & 2     & 1     & 0         & 0     &  19     & 65    & \multicolumn{1}{c}{} \\
          & 1     & 0     & 100     & 100   &   0    & 0     & \multicolumn{1}{c}{} \\
          & 0     & 0     & 0          & 0     &  0     & 0     & \multicolumn{1}{c}{} \\
\cline{1-7}    \multicolumn{1}{r|}{\multirow{4}[2]{*}{Fair}} & 3     &\textbf{ 97}    & 0         & 0     &    71   & 12    & \multicolumn{1}{c}{} \\
          & 2     & 3     & 0         & 2     &   19    & 81    & \multicolumn{1}{c}{} \\
          & 1     & 0     & 100   & 98    &   0    & 7     & \multicolumn{1}{c}{} \\
          & 0     & 0     & 0        & 0     &    0   & 0     & \multicolumn{1}{c}{} \\
\cline{1-7}    \multicolumn{1}{r|}{\multirow{4}[2]{*}{Low}} & 3     & \textbf{80}    & 0         & 0     & 40    & 1     & \multicolumn{1}{c}{} \\
          & 2     & 19    & 4     & 2     & 60    & 52    & \multicolumn{1}{c}{} \\
          & 1     & 1     & 96        & 98    & 0     & 47    & \multicolumn{1}{c}{} \\
          & 0     & 0     & 0         & 0     & 0     & 0     & \multicolumn{1}{c}{} \\
       \hline
\multicolumn{1}{c}{}&\multicolumn{1}{c}{}&&&&&&\\
\multicolumn{1}{c}{}&\multicolumn{7}{c}{\textbf{(b) Number of noise variables selected}}\\
    \hline
  \multicolumn{1}{r|}{High} &\multicolumn{1}{c}{mean}&\textbf{0.040} & 0.360   & 0.080   &15.16 & 52.540 & \multicolumn{1}{c}{\multirow{3}[2]{*}{}} \\
   \multicolumn{1}{r|}{} &\multicolumn{1}{c}{SD}&(0.197) &  (0.503)  &  (0.273) &  (4.334) &(29.186) & \multicolumn{1}{c}{\multirow{3}[2]{*}{N/A}} \\
\cline{1-7} 
  \multicolumn{1}{r|}{Fair} &\multicolumn{1}{c}{mean}&0.140 &0.790 &\textbf{0.100}  &  4.436&46.370 & \multicolumn{1}{c}{} \\
   \multicolumn{1}{r|}{}&\multicolumn{1}{c}{SD}& (0.377) &  (0.686) & (0.302) & (2.106) & (30.833) & \multicolumn{1}{c}{} \\
\cline{1-7} 
  \multicolumn{1}{r|}{Low}  &\multicolumn{1}{c}{mean} & 0.540 &0.700 & \textbf{0.010}& 14.330  & 28.700 & \multicolumn{1}{c}{} \\
   \multicolumn{1}{r|}{}  & \multicolumn{1}{c}{SD}& (0.784) & (0.916)  &  (0.100) &  (5.650) &  (16.407) & \multicolumn{1}{c}{} \\
  \hline
\multicolumn{1}{c}{}&\multicolumn{1}{c}{}&&&&&&\\
&\multicolumn{1}{c}{Oracle MSE}&\multicolumn{7}{c}{\textbf{(c) Median MSE (IQR) in test samples}}\\
   \hline
   \multicolumn{1}{r|}{High} &\multicolumn{1}{c}{1}&   \textbf{1.009} &  4.433  &  4.555  & 2.781&  3.470  &2.977 \\
   \multicolumn{1}{r|}{} &\multicolumn{1}{c}{} &(0.131)&(0.672)  &(0.678) &   (0.379)    &  (0.794) &  (0.416)   \\
\hline
   \multicolumn{1}{r|}{Fair} &\multicolumn{1}{c}{1}&  \textbf{1.004}  &2.870 &   3.117 &2.083&  2.535 & 2.236 \\
   \multicolumn{1}{r|}{} &\multicolumn{1}{c}{}&  (0.143) &  (0.430)  &(0.357) &   (0.258)    & (0.460) &  (0.272) \\
  \hline
   \multicolumn{1}{r|}{ Low}  &\multicolumn{1}{c}{1} &  \textbf{1.011}  & 1.691   & 1.711& 1.428  & 1.644  & 1.431 \\
   \multicolumn{1}{r|}{} &\multicolumn{1}{c}{} &  (0.176)& (0.267)  &  (0.237) &(0.232) & (0.499) &  (0.224)   \\
   \hline

\multicolumn{1}{c}{}&\multicolumn{1}{c}{}&&&&&&\\
&\multicolumn{1}{c}{Average Oracle $R^2$  (sd)}&\multicolumn{7}{c}{ \bf{(d) $R^2$} in test samples}\\
   \hline
      \multicolumn{1}{r|}{High} &\multicolumn{1}{c}{0.803}&\textbf{0.801} &  0.367  &  0.356  & 0.460& 0.324  &0.417 \\
   \multicolumn{1}{r|}{} & \multicolumn{1}{c}{ (0.018)}&(0.018)&(0.044) &(0.046) &     (0.048)  &  (0.109) &(0.038)  \\
   \hline
   \multicolumn{1}{r|}{Fair} &\multicolumn{1}{c}{0.749 }& \textbf{0.706}  &0.389 &   0.336&0.391&0.267   &0.359 \\
   \multicolumn{1}{r|}{} &\multicolumn{1}{c}{(0.029)}&  (0.030) &  (0.072)  &(0.050) &  (0.045)     & (0.100) &   (0.037)  \\
   \hline
   \multicolumn{1}{r|}{ Low}  & \multicolumn{1}{c}{0.455}&   \textbf{0.439}  & 0.274  &0.253& 0.215  &0.127  &0.211  \\
   \multicolumn{1}{r|}{ }  &\multicolumn{1}{c}{ (0.051)} &  (0.051)& (0.070) &  (0.054) &(0.062) & (0.071) &  (0.053)   \\
   \hline
\multicolumn{1}{c}{}&\multicolumn{1}{c}{}&&&&&&\\
    \end{tabular}%

  \label{tab3}%

\end{threeparttable}%
  }
\end{table}

\subsection{Study 3: Simulation with more true predictors}

We also performed a more challenging simulation with 8 true
predictors, covering a range of taxonomic levels and rareties of taxa,
also with different signal strengths for different taxa. Full details
of the simulation are presented in the online supplementary materials.
As expected, variables with larger coefficients are more easily
selected. However, rarer taxa are selected less often, even when they
have relatively high coefficients. These patterns are common to all
variable selection methods. Across the range of signal-noise ratios
and coefficients, SuRF outperforms stability selection (with default
family wise error rate upper bound) in terms of both false positive
and false negative rate. SuRF hugely outperforms LASSO in terms of
false positive rate, and outperforms LASSO in false negative rate at
high SNR, with comparable performance at lower SNR. In terms of
misclassification error rate, SuRF is clearly the best method. We did
not compare VSURF in this simulation because of its slow running time.

\section{Application: the pouchitis and moving picture data}
Two published datasets are analyzed using SuRF. The first dataset is
from a pouchitis study \citep*{tyler2013characterization}.  The second
dataset inlcudes samples from four body sites of two individuals over
a long time period \citep*{caporaso2011moving}.

\subsection{Pouchitis study}

Colectomy with ileal pouch anal anastomosis (IPAA), also
referred to as ``J-pouch surgery'', is a common surgery for patients
who have ulcerative colitis (UC) and those with familial adenomatous
polyposis syndrome (FAP) \citep*{Shen2013Pouchitis}. Pouchitis is a
common complication of J-pouch surgery involving inflammation of the
ileal pouch. It is unclear what triggers pouchitis in some patients
but not others: pouchitis occurs almost exclusively in patients with
inflammatory bowel disease and not in patients with FAP.

Our data come from the study \citet*{tyler2013characterization} which
includes microbiome samples from biopses of 71 patients following a
J-pouch surgery. Our objective is to classify individuals between the
healthy and inflammation group. The inflammation group is composed of
the 34 subjects from the ``pouchitis'' and ``CD (Crohn's
disease)-like'' groups in the original paper. It includes inflammation
in either the pouch or the pre-pouch ileum; and the inflammation may
or may not be active at time of biopsy. The healthy group is composed
of the 37 subjects in the ``FAP'' and ``no pouchitis'' groups from the
original study.

Some patients received one or two antibiotic treatments before the
biopsy. We include two variables describing antibiotics
usage in addition to the proportions of OTUs at each taxonomic rank,
making a total of 1781 predictors. The same information was
measured at both pouch and afferent limb for each patient.

The mean classification error rate is estimated by averaging the
cross-validated classification error across a thousand subsamples. It
is about 0.2 and 0.35 for pouch and afferent limb, respectively.  At
both biopsy sites, the phylum Bacteroidetes is the only variable
significant at level 0.05. The agreement on the importance of
Bacterioidetes at both biopsy sites suggests this phylum is
significantly associated with inflammation. The single bacteriodetes
phylum gives a 0.88 and 0.83 AUC (area under the ROC curve shown in
Web Figure~3) in the pouch and afferent limb respectively. The ROC
curves suggest that bacteriodetes is an effective discriminant
variable for differentiating the inflammation condition at both biopsy
sites, especially for the pouch data.

Even the non-significant highly ranked variables are potentially interesting variables for future studies.  Among the top variables in both sites, there are
several common variables, which are Bacteroidetes, 
Bacteroidia, 
Erysipelotrichi,
and Bacilli. 
Most of these organisms have been found to be
associated with IBD and pouchitis in the literature. Bacteroidia is a major
class of the Bacteroidetes phylum. 99.7\% of this phylum are
Bacteroidia in this dataset. Both taxonomy levels are found to be
negatively associated with the disease at both biopsy sites. This is
consistent with previous findings from other datasets that there is a
decreased abundance and diversity of Bacteroidetes in CD samples
\citep*{de2015microbiota}. Similar findings are also reported in
\citet*{tyler2013characterization} using the same dataset we have
analysed here. They identified that Bacteroidetes are significantly
reduced in the pouchitis and CD-like groups compared to the FAP and no
pouchitis groups. Many of the other highly ranked OTUs
have been linked with related conditions such as IBD in previous
studies.  The family Fusobacteriaceae has previously been found to
have a higher proportion in patients with UC who underwent pouch
surgery \citep*{reshef2015pouch}. 
Though we know little about the role of \textit{Turicibacter} in IBD
\citep*{jones2015ablation}, it has been shown significantly decreased
in dogs with IBD \citep*{rossi2014comparison}.  Bacilli and
Erysipelotrichi have been found increased in patients with UC
\citep*{michail2012alterations} and low counts of
\textit{Subdoligranulum} have been found in patients with Crohn's
disease \citep*{thomas2014exploring}.

Stability selection also selects Bacteroidetes for cut-off
probability 0.6 for both pouch and afferent limb, but selects no
variables at higher cut-off probabilities 0.8 and 0.9. At cut-off
probability 0.7, it selects Bacteroidetes for the pouch data, but
selects no variables for the afferent limb data.
In Table~\ref{tab6}, we compare the predictive accuracy of the logistic
regression model using the selected variable Bacteroidetes, with other
commonly used classification methods for microbiome data, namely
Random Forest (RF) and Support Vector Machine (SVM) with a linear
kernel (we obtained similar results for other kernels and omited them
from the table). These predictive accuracies are computed using
leave-one-out cross-validation with their corresponding tuning
parameters chosen by cross-validation within the training data. The
predictive accuracy from RF and SVM are comparable to the results
using SuRF since the mean test errors are all within one
standard deviation.


\subsection{Moving picture data}

The moving picture data set \citep*{caporaso2011moving} recorded a
long period of repeated observations from multiple body sites (gut,
tongue, left and right palms) of two individuals. 
This data set has a larger sample size for each body site than the
pouchitis data.  
The number of observations for the gut, tongue, left palm and right palm are respectively 
131, 135, 134 and 134 for the first individual, and 
336, 373, 365 and 359 for the second individual. We split the dataset for
each site into a training and a test sample set with a ratio of
2:1. At each body site, the observations from each individual are
ordered by time. The earlier 2/3 of time points from each individual
are used as training samples and the rest as test samples.

We train SuRF to classify samples from each body site between the two
individuals using the training data. The selected variables are
summarised in Web Table~4, and the joint distributions are displayed
in Web Figure~4. Table~\ref{tab6}(b) shows the misclassification error
rate for SuRF and othe methods. Between one and four variables are
selected at each body site and the prediction errors for the test
samples are very low at all sites. SuRF has found a small set of
variables that can distinguish two individuals' microbial
environments. For most methods the test error tends to be lowest in
the gut and highest for palms. This can be well explained by the fact
that the microbiome community is most stable in the gut
\citep*{voigt2015temporal} and least stable for palms because, in
contrast to the human gut, the composition of microbial communities
from hands, though in the long run relatively stable
\citep*{oh2016temporal} and personalized \citep*{fierer2010forensic},
can change dramatically 
even from 
washing hands with some disinfectant cleaning products. 
Identifying individuals using the palm microbiomes
is feasible but more variabile than using
a more closed environment such as the gut.

We also tested cross-predictions --- using models fitted on one body
part to predict the owner of samples from another body part. The
prediction model trained on one palm could identify samples from the
other palm with low prediction error. The two bacteria selected in the
two palm models include the same species-level variable from genus
\textit{Deinococcus} and two different unspecified species-level
variables from genus \textit{Corynebacterium}. This suggests a
similarity between the microbiomes on two palms from a single
individual. No other cross-predictions performed
significantly better than random guessing.

SuRF and stability selection (using cutoff probability 0.9 and default
family error upper bound) were on average comparable in predictive
accuracy to Random Forests and significantly better than SVM (see
Table~\ref{tab6}(b)).  Compared to stability selection, we found that
SuRF seemed to achieve a lower prediction error, and consistently
selected fewer variables.

In the gut data, SuRF chooses one unspecified species from the genus
\textit{Bacteroides}, which is one of three variables selected by
stability selection with cut-off probability 0.9.  With one variable
we obtain exactly the same prediction training and test errors as with
three variables selected by stability selection.  The other two
variables selected by stability selection (another unclassified
species from the genus \textit{Bacteroides} and the family
Porphyromonadaceae) don't provide additional predictive accuracy for
recognizing individuals.

In the tongue data, even using cut-off probability 0.9, stability
selection still selects eight variables. SuRF selects only three
variables: the most important variable is one species from genus
\textit{Neisseria} and the remaining two are unspecified species from
the family Lachnospiraceae, and order Sphingobacteriales. There are no
common variables selected by both SuRF and stability selection. This
is the case where SuRF performed less well than other methods, so it
is natural to ask whether SuRF might have selected too few variables
in this case. However, using only the first two variables chosen by
SuRF reduces the test error to 0.03, so the poor performance here is
not entirely explained by excessive sparsity.

For the left palm data, both stability selection with the highest
cut-off probability and SuRF choose the same set of variables (one
unspecified species from the genus \textit{Corynebacterium} and another
unspecified species from \textit{Deinococcus}). 

For the right palm data, SuRF selects the same species from the genus
\textit{Deinococcus} and a different unspecified species from the
genus \textit{Corynebacterium}. The former is also selected by
stability selection for the right palm model, but the second variable
is replaced by the kingdom Bacteria. Both methods choose two variables
(using cut-off 0.9 for stability selection), however, SuRF not only
provides a smaller prediction error for both training and test data,
but also indicates a similarity between two palms within the
individual which is not reflected by the variables selected by
stability selection.

These two real datasets exemplify the ability of SuRF to select
discriminant OTUs at the appropriate taxonomic level. For the
pouchitis data, with large within-class variation at lower levels,
SuRF identifies a phylum-level variable. For comparing two healthy
individuals, the higher-level structure is more
similar, so SuRF selects species-level variables.

\begin{table}[!htbp]
 \resizebox{0.9\textwidth}{!}{%
\begin{threeparttable}[t]

  \centering
  \caption{Results comparison among
    SuRF, Stability selection, VSURF, LASSO, Random Forest (RF) and SVM (Linear Kernel) for the pouchitis study and moving picture data}

    \begin{tabular}{l|cc|ccc|cc|cc|c|c}
    \hline
 \hline
        \multicolumn{1}{l|}{Data} & \multicolumn{2}{c|}{SuRF} &        \multicolumn{3}{c|}{Stability Selection} & \multicolumn{2}{c|}{VSURF} & \multicolumn{2}{c|}{LASSO} & RF &SVM  \\    
  \hline
\multicolumn{12}{l}{\bf{(a) Mean Misclassification error (sd) }}\\
 \hline
    Pouch\tnote{1} & \multicolumn{2}{c|}{0.197 (0.047)}   &  \multicolumn{3}{c|}{0.197 (0.047)}   & \multicolumn{2}{c|}{0.268 (0.053)}  & \multicolumn{2}{c|}{0.282 (0.053)}  &\textbf{0.169} (0.044) &0.211 (0.048)  \\
 Afferent limb\tnote{1}& \multicolumn{2}{c|}{0.254 (0.052)}   &  \multicolumn{3}{c|}{0.254 (0.052)}   & \multicolumn{2}{c|}{0.254 (0.052)}  & \multicolumn{2}{c|}{0.324 (0.056)}  &0.225 (0.050) &\textbf{0.211} (0.048)  \\
  Gut & \multicolumn{2}{c|}{0.000}   &  \multicolumn{3}{c|}{0.000}   & \multicolumn{2}{c|}{0.000}  & \multicolumn{2}{c|}{0.000}  &0.000 &0.000 \\
 Tongue& \multicolumn{2}{c|}{0.053 (0.017)}   &  \multicolumn{3}{c|}{\textbf{0.000} }  & \multicolumn{2}{c|}{0.018 (0.010)}  & \multicolumn{2}{c|}{\textbf{0.000}}  &0.006 (0.006) &0.024 (0.012)  \\
    Left Palm & \multicolumn{2}{c|}{\textbf{0.024} (0.012)}   &  \multicolumn{3}{c|}{0.030 (0.013)}   & \multicolumn{2}{c|}{0.061 (0.05319)}  & \multicolumn{2}{c|}{0.079 (0.021)}  &0.079 (0.021) &0.224 (0.032)  \\
 Right Palm& \multicolumn{2}{c|}{\textbf{0.025} (0.012)}   &  \multicolumn{3}{c|}{0.067 (0.020)}   & \multicolumn{2}{c|}{\textbf{0.025} (0.012)}  & \multicolumn{2}{c|}{0.129 (0.026)}  &0.037 (0.015) &0.288 (0.035)  \\
 Left predict Right Palm& \multicolumn{2}{c|}{0.020 (0.006)}   &  \multicolumn{3}{c|}{0.020 (0.006)}   & \multicolumn{2}{c|}{\textbf{0.014} (0.005)}  & \multicolumn{2}{c|}{0.049 (0.010)}  &0.152 (0.016) &0.148 (0.016)  \\   

   \hline

    \multicolumn{12}{l}{\bf{(b) The total number of variables selected }}\\
\hline
Pouch\tnote{1} & \multicolumn{2}{c|}{\textbf{1 (0)}}   &  \multicolumn{3}{c|}{\textbf{1 (0)}}   & \multicolumn{2}{c|}{4.282 (0.701)}  & \multicolumn{2}{c|}{1.254 (1.795)}  &\multicolumn{2}{c}{\multirow{6}[2]{*}{N/A}}  \\
Afferent limb\tnote{1}& \multicolumn{2}{c|}{\textbf{1 (0)}}   &  \multicolumn{3}{c|}{\textbf{1 (0) }}  & \multicolumn{2}{c|}{6.592 (0.729)}  & \multicolumn{2}{c|}{2.676 (12.033)} &&   \\
 Gut & \multicolumn{2}{c|}{\textbf{1}}   &  \multicolumn{3}{c|}{3}   & \multicolumn{2}{c|}{\textbf{1}}  & \multicolumn{2}{c|}{18}  &&  \\
 Tongue& \multicolumn{2}{c|}{\textbf{3}}   &  \multicolumn{3}{c|}{8 }  & \multicolumn{2}{c|}{\textbf{3}}  & \multicolumn{2}{c|}{9} &&   \\
    Left Palm & \multicolumn{2}{c|}{\textbf{2}}   &  \multicolumn{3}{c|}{\textbf{2}}   & \multicolumn{2}{c|}{3}  & \multicolumn{2}{c|}{67} && \\
 Right Palm& \multicolumn{2}{c|}{\textbf{2}}   &  \multicolumn{3}{c|}{\textbf{2}}   & \multicolumn{2}{c|}{4}  & \multicolumn{2}{c|}{45}&&   \\

   \hline

    \end{tabular}%
    
  \label{tab6}%
\begin{tablenotes}
\item [1] Misclassification error and mean number of variables selected with standard deviation are cacluated based on leave-one-out  for pouch and afferent limb.\\
\end{tablenotes}

\end{threeparttable}%
  }

\end{table}

\section{Concluding Remarks} \label{concluding remarks}

We have developed a very useful variable selection method for GLMs,
SuRF, which involves a subsampling based approach to rank variables
that may be highly associated with the response variable followed by
variable selection with forward ANOVA. This method takes advantage of
the sparseness of the model selected by LASSO and chooses variables
that appear more frequently and contribute significantly to reducing
residual deviances in the forward ANOVA procedure. Due to its high
sparseness and stability, SuRF can be particularly useful for
microbiome data or any data that is high dimensional and contains many
surrogate variables. The method provides a conservative but stable
selection of variables that can predict and classify the
outcomes. SuRF can also provide a reasonable way to compute $p$-values
for all variables according to sequentially calculated empirical
distributions, whereas LASSO does not provide $p$-values directly.
The forward selection procedure helps to alleviate the phenomenon of
including surrogate variables and leads to a highly sparse model for
microbiome data. Due to its short list of selected variables SuRF is
particularly suitable for identifying biomarkers.

In our simulation studies we saw that in comparison to many competing
methods, SuRF is able both to select the true variable more often, and
also to select fewer noise variables for both binary and continuous
outcomes. This leads to excellent performance in prediction.

In the two real data analyses, we found that no other methods
significantly outperformed SuRF in terms of prediction error, but SuRF
selects fewer variables than other methods. SuRF was able to adjust the
taxonomic levels of the variables selected to suit the individual datasets.

There are many promising avenues for future research into extending
the SuRF framework. In this paper, we have presented SuRF based on
penalised regression followed by generalised linear models, because
that seemed most appropriate to the structure of the microbiome
data. However, the core idea is to use subsampling with a simple
variable selection method, then use the ensuing ranking in a forward
selection method. This core idea could be applied with any combination
of a variable selection method and a family of nested models to be
used in forward selection. For example, we could develop a ranking
based on Random Forest, and then perform the forward selection based
on neural networks. The use of the permutation test for evaluating a
variable automatically adjusts to our choice of method. Further
research is needed into what combinations of methods work well in this
framework.

\section{Supplementary Materials}
Web Appendices, Tables, and Figures referenced in Sections 2--4 are available with this paper at the Biometrics website on Wiley Online Library.

An \texttt{R} package for applying SuRF is available from Toby~Kenney's
website at \url{www.mathstat.dal.ca/~tkenney/Rpackages/}. The authors
also plan to submit this package to CRAN.
The package uses the \texttt{glmnet} package for fitting LASSO
regression, and can be used for any models that the \texttt{glmnet}
package supports.




\label{lastpage}

\end{document}